\documentclass[floatfix, twocolumn, preprintnumbers,amsmath,amssymb,prb,superscriptaddress]{revtex4}
\usepackage{amsmath}
\usepackage{graphicx}%
\usepackage{dcolumn}
\usepackage{amsmath}
\usepackage{subfigure,hyperref}
\usepackage[normalem,normalbf]{ulem}

\usepackage{enumerate}

\newcommand{\be}{\begin{equation}}
\newcommand{\ee}{\end{equation}}

\usepackage{enumerate}

\begin{document}

\title{No such thing as a risk-neutral market }
\author{D.L. Wilcox}
\email{diane.wilcox@wits.ac.za}
\affiliation{School of Computer Science and Applied Mathematics, University of the Witwatersrand, Johannesburg, South Africa}
\affiliation{ QuERILab - Quantifying Emergence, Risk and Information}

\begin{abstract}

A very brief history of relative valuation in neoclassical  finance since 1973 is presented, with attention to core currency issues for emerging economies.   Price formation is considered in the context of hierarchical causality, with discussion focussed on identifying mathematical modelling challenges for robust and transparent regulation of interactions.

\vspace{0.3cm} \hspace{-0.5cm}

\end{abstract}

\maketitle


In order to illustrate the complex interplay between derivative markets and underlying economies, this essay includes an abridged record of some key determinants of currency valuation.
 Arguments are qualitative rather than quantitative and aim to highlight the need for better representation of market information and for regulation to ensure  pricing  in developing economies is protected from systemic arbitrage.

 Attention is given to  some  repercussions of the supply and transfer of money and capital across borders from an emerging market perspective, specifically South Africa.
  While  this note does not address money supply within any country or legacy capital ownership in a post-colonial era, it identifies  a few  geopolitical forces in generality. In particular, given the historic links between SA capital markets and UK economic interests, as well as the continued global dominance of the US economy in the post cold war era, some hierarchical impacts of their global policies on risk-neutrality are considered.

 Overall, the discussion aims to give insight into the multilevel modeling of a key economic variable, taking into account endogenous and exogenous sources of causation.

\vspace{-0.2cm}

\section{The advent of risk-neutral pricing and dynamic security hedging}

\vspace{-0.2cm}

Derivative pricing entails the hedging of market risks. The breakthrough of  the Black-Scholes-Merton (BSM) model was the use of extremely elegant mathematical formalism to show that risk could be eliminated under simplified assumptions.  The expansion of trade in derivative instruments has been documented from numerous perspectives: notional amounts in global markets now aggregate to  hundreds of trillions of US dollars,  low points include the LTCM crisis (1998) and failure of Lehman Brothers  (2008) and extreme points include the hundreds of billions in quantitative easing (QE) to alleviate the crisis caused by defaulting credit risk contracts under limited global regulatory oversight (2008-present).

A characterisation of no-arbitrage markets emerged round the same time as the BSM derivative pricing model: commencing with a simplified finite-state model, Stephen Ross provided no-arbitrage conditions in the so-called first fundamental theorem of asset pricing.  This ensured the existence of a consistent pricing mechanism, via a risk-neutral measure for expected prices, which does not allow a ``free lunch'', i.e. attainment of a riskless profit. 

Equivalently, if two products are effectively identical (in every measurable sense), then they  should cost the same.  More specifically, if the net cash-flows which they generate over the lifetime of the contracts are equivalent, then two investments are considered to be equivalent. If either is mispriced  and all else is equal, then speculators would step in to exploit the difference by purchasing the cheaper version and selling it at the higher price, making profit without holding inventory.

No arbitrage is closely coupled to the notion of market efficiency, whereby all assets are priced according to correct information which is available instantaneously to all market participants. A routine shopping exercise  to discern between toothpastes sold in a large supply store highlights that efficient decision making is constrained by the ability to reason and weigh a wide range of benefits and costs in reasonable time, possibly ignoring useful detail. Conditions in illiquid markets typically imply that transaction costs are higher, where additional costs are incurred to fund  due-diligence and reliable information gathering.

When the BSM papers were published in 1973,  asset valuation under fiat currency was premised on the capacity to understand how different assets store and produce value over different time horizons and hence,  to assess correct discount and inflation rates and keep markets honest.  In theory, novel exchange-traded futures and options contracts enabled lower risk costs for managing uncertainty.
 
 The notion of ``no free lunch'' has been generalised to a more sophisticated mathematical formalism for consistent pricing of investment portfolios, based on the non-admissability of arbitrage trading strategies and referred to as ``no free lunch with vanishing risk''. Thus, risk-neutral pricing of tradeable assets offered a theoretic framework which made aggregate market growth consistent with the supply of capital through monetary policy.

\vspace{-0.5cm}

\section{No such thing as a free market}

\vspace{-0.2cm}

Free market economist Milton Friedman published ``There's no such thing as a free lunch'' in 1975.  The title echoed a well-used phrase  from the real US economy of the 1930's.
Friedman was interested in removing all regulatory constraints while advocating that risk had a cost which market forces could be relied on to anticipate correctly in natural pricing. Coupled with a view that wealth trickled down into the real economy, liquid derivative markets were considered as a path to making the pricing processes for stocks, bonds and commodities more efficient.

Geopolitics of the post Bretton Woods era was far more complicated than any simplified model. In reality, US economic policy makers were confronted with a market-changing petrodollar-shock delivered by OPEC while exiting it occupation of  Vietnam. The beginning of the 1970's also saw support for popular socialist movements around the world,  with democratic elections voting into power  local control of wealth in the interest of local communities.   However, it was still the cold world era and NATO continued anti-communist interventions in the interest of global capital.

As multinational corporations took up residences around the world for new markets, cheaper labour and tax arbitrage through the 1980's, US economic policy from Washington advocated for  so-called open markets.  Simultaneously, subsidies and rebates supported its own suppliers of energy and agriculture goods.

By the turn of the millennium, members of the US Federal Reserve banks and the City of London wrapped up Reagonomics and Thatcherism by removing the last  market constraints which had been set in place to contain moral hazard after the 1929 crash.  Deregulation of major capital markets provided lower-cost credit for the debt-funded growth of  stake-holders.

Market crashes are not the worst failures which an economy can suffer. If one considers  price formation in its simplest form,  a buyer and seller meet and exchange bids and offers until they converge on an agreeable price or walk away from the auction. Unnatural market crises occur when  failed auctions lead to one party assaulting the other to demand a price.

Former US Federal Reserve bank governor, Alan Greenspan, was famous for serving under a long duration of market growth which is referred to as the great moderation (1982-2007). At its high point, it was advocated that US economic policy had successfully tamed the management of market crises. However, his successor's acknowledgement of the successful diminishing of market volatility in 2004 came a year after the US invasion of Iraq. Since then, Greenspan  has been quoted in leading mainstream media giants to say  that the military occupation of Iraq, which was imposed without sanction by the UN,   was indeed about oil interests. In reality, the  taming of profit-seeking allocation of free capital  coincided with the perpetuation of the doctrine reiterated in 1980 by Carter that the US would use military force if necessary to defend its national interests in the Persian Gulf.

There are grave economic implications when dominant economies are backed by the biggest military-industrial complex and do not follow  globally accepted procedure. Given US protection of some very restricted societies, it is consistent that oil and energy interests drive the highly selective nature of US protection of universal franchise and global openness. Such distortions can have permanent impact on  the natural evolution of no-arbitrage conditions.

At the WEF session on the Global Economic Outlook at the start of the 2016, UK Chancellor of the Exchequer remarked that ``the world has not very been good $\ldots $ at accommodating rising powers.''
China currently  holds approximately US\$ 1.3 trillion of US treasuries or, equivalently, 10$\%$  of US National debt, which stands at about \$US 13.5 trillion. It follows that the inclusion of the Yuen as an IMF reserve currency is consistent with its persistent stake in US monetary supply, purchased with the output of decades of labour in market driven production for global consumers.

\vspace{-0.2cm}

\section{Co-existing currencies and risk asymmetry}

\vspace{-0.2cm}

While cheap oil and consumption refueled markets after the NASDAQ crash of 2000, the great moderation ended abruptly with the onset of the global financial crisis. Documentations of the free market failings which led to the dotcom bubble  and the global financial crisis (GFC, 2007+) are manifold, with numerous  bestsellers  published to film and print media. Greenspan himself has admitted that the Washington consensus had gotten its models for moderation wrong.

The contagion effects of market mispricing of credit derivatives were global, with economies like SA markets rocked dramatically without significant direct investment in the defaulting assets. Interventions by  regulators to address the crisis fallout generated significant secondary impacts on  developing markets.

As members of smaller economies reassessed their purchasing power, hundreds of billions of USD,  EUR,  GBP and YEN  were freed onto the global markets under quantitative easing. Policies to bolster developed markets out of recession ignored potential knock-on effects of renewed speculative investment in volatile global markets.  Instead of   simply alleviating debt crises in the intended target markets, the unregulated channeling of QE capital into emerging markets contributed to the ad hoc risk of  withdrawals under uncertain allocations of further rounds of QE, as well as devaluation challenges.

With  some estimates for the capitalisation of non-bank financial intermediaries in the range of US$\$$ 45 - 75 trillion, major regulators have acknowledged the need to address the systemic impact of so-called shadow banking.    Defaulting off-balance-sheets contracts at the heart of the credit crisis were moved onto the balance sheet of the  US Federal Reserve bank  through QE. However, even though it has increased its balance sheet since 2008 to US\$ 4.5 trillion, the US Fed Reserve bank is dwarfed by the potential uncertainty of disruption from non-bank capital infrastructure.

Significant  unintended consequences of regulation in SA came  under Apartheid itself. While affluent South Africans enjoyed  the largest part of wage share and capital accumulation in the 60's and 70's, higher GDP was not able to tame  anxiety over ongoing threats of ``terrorist'' defiance from the disenfranchised. The  financial rand as a parallel currency was first introduced in the 1960's, dropped in the early 80's and then reintroduced in 1985, in order to regulate the tide of locally-produced capital leaking outward through currency boundaries. Unlike the fabled finger in the dyke, this strategy failed to  stem outward flow. Despite the monetary interventions to stimulate investment in SA assets, the Apartheid economy had unsustainable in both real and financial ZAR terms.

Deliberate misinformation corrodes market quality further.  In ``Phishing for Phools'', Shiller and Akerlof warn of the moral hazard in unregulated markets from an endogenous perspective, whereby competition drives unscrupulous market participants to cheat. Even more cynically, misinformation can become an effective tool for price manipulation at global scale for participants with enough leverage, as borne out by the mispricing of risk and long-horizon cashflows.

 Litigation of the malpractices in issuance and trade of debt, which resulted in  overinflated US house prices,  debt hidden in shadow credit vehicles and destabilized global markets,  has led to banking fines of more than US \$200 billion. Given the corresponding credit relief to those same institutions and subsequent waves of QE, the credit crunch and aftermath have  highlighted the non-homogeneous impact of regulatory influence and financial innovation in complex hierarchies.

The existence of large-scale arbitrage opportunities imply that markets are not risk-neutral in the sense of Ross's neoclassical finance. In particular, this implies that it is impossible to value future global cashflows consistently over extended periods, even with robust models for the underlying rates.

\vspace{-0.2cm}

\section{Modelling the multifaceted nature of relative monetary flows}

\vspace{-0.2cm}

The NYSE market crash of Black Monday of 1987 offers  one of the first global failures of the  application of dynamic hedging strategies. Since then the BSM model has been revised many times over  to incorporate better noise analysis, additional underlying variables for credit and liquidity risks and some cost for model uncertainty.

Today, advanced mathematical pricing models continue to provide trading strategies for hedging cashflows in more than one currency and taking into account coupled dynamics for credit risky payments. Case studies for emerging markets include scenarios whereby bond default of a large parastatal could trigger foreign-exchange devaluation and conversely, currency deterioration could cause a credit crisis for company or portfolio which needs to deliver foreign-denominated payments.

At exchange interfaces, information transmission and price evolution are not free from process errors on trading platforms. In this domain, decades of research have been allocated to the study of order flows, to the extent that short range prediction as well as anomaly detection are possible.

Given the multifaceted nature of value and exchange currency for trade within a single closed economy, it is clear that exchange rates are impacted by almost all economic variables, directly or indirectly. Any attempt at predictive modeling is forced to grapple with dimension reduction of a highly complex system.

While many data mining methods to analyse exchange rates  are able to offer short range predictions, they do not necessarily provide comprehensive descriptions of causality. At the opposite end, the sort of  optimal foreign exposure hedge ratios provided by financial economists like Fisher Black relied on idealised input variables for investment between perfect markets.

Research on cross-border flows is already mature in the sense that much has been written and empirical approaches to analysing  currency are based on  a wide spectrum of investment and consumption data. Early attempts to aggregate data include so-called gravity models, named  for their dependence on variations of size of and estimations of square-distances between economies. However, such approaches to obtain single equations to describe price relationships are bounded by their simplified dependence on underlying variables.

A variety of models have been developed to analyse portfolios of currencies. Covariance analysis has been used to investigate co-movements of exchange-rates and hierarchical dependencies, mostly between developed markets. While these analyses typically ignore other economic variables,  the approach is able to map the evolution of  dominant clusters of  causation directly, giving a summarised perspective of trade relations.

Even in complex economies such as SA,  asset-management has defended market quality by appealing to the belief that market forces are  most efficient with respect to capital allocation.  Policies for attracting foreign direct investment (FDI) and net portfolio investment (NPI)  were promoted with the implicit
assumption that benefits would reach required areas for development. From an abstract  mathematical modelling perspective, insight from the study of dynamical systems has exposed how sensitivity to initial conditions can result in dramatically varying evolutions or, equivalently, that there is an inescapable flaw in assuming that an uncontrolled system will reach homogeneous or equitable conditions, irrespective of initial and boundary conditions.

For emerging markets, investigations into macroeconomic  determinants of  FDI and NPI  have included regression analysis against  price stability, stable policies,  transparency, openness to trade, infrastructure and lack of corruption. However,  empirical evidence has also exposed   that the most vulnerable, small-capitalised companies in a market can be impacted the most under changes in FDI, even when there exist favourable investment conditions in the target economy.

Given that global power blocks engaged in closed market negotiations and military interventions in developing markets during the cold-war era,  foreign investment is inextricably linked to geopolitical forces. Thus,  there are numerous caveats associated with the notion of openness championed by neo-liberalism and regulation is required to ensure that investment benefits include local infrastructure development for the public good.

By some measures, China can now be regarded as the biggest national economy on the global trade field. At the same time other G20 and BRICs economies are at still different stages of development relative to the G7. Yuen re-valuation and a shift in Chinese economic focus from infrastructure to consumption have implications for the demand for resources from South Africa. On the other hand FDI from China into other emerging economies offers new potential, provided such investments drive development in the real economy.

Many currency paradigms have ignored debt. While countries like SA were held accountable for repaying Apartheid government debt even after democratisation in 1994, management of private and government debt  in countries such as the US are regulated by different rules, with trillion dollar debt a source of uncertainty for emerging markets even as global policy makers intervene to ensure global stability. Despite significantly lower debt to GDP ratios, global market sentiment advocates austerity over QE for  emerging markets to ensure future prosperity.

 Deep differences exist with respect to how economists explain the role of debt and money supply in the GFC. Steve Keen argues that the omission of the dynamics of private debt  is the key failure of the macro-economic modelling which culminated in systemic failure.
His approach addresses debt as endogenous money to deduce that change in money supply is equivalent to change in debt. In contrast some neoclassical perspectives equate debt to loanable funds with zero impact on net money.

Following the GFC, there are still unresolved questions for consistent accounting of  cashflows in the valuation of derivative contracts. With debt-related payments coupled to adjustments for bilateral credit risks measurements,  ongoing challenges include how to mitigate against default in networks of non-bank liabilities and implications of rehypothecation in the case of collateralisation.

The models in this section emerge from differing perspectives to provide insight on both  endogenous evolution and  impacts of exogenous changes. If one zooms into the challenge of currency or exchange rate valuation, it is clear that required methodologies are far more advanced than when interactions were considered in 1973. A non-exhaustive list of determinants of local currency value would include:

\begin{figure}
    \includegraphics[width=9cm]{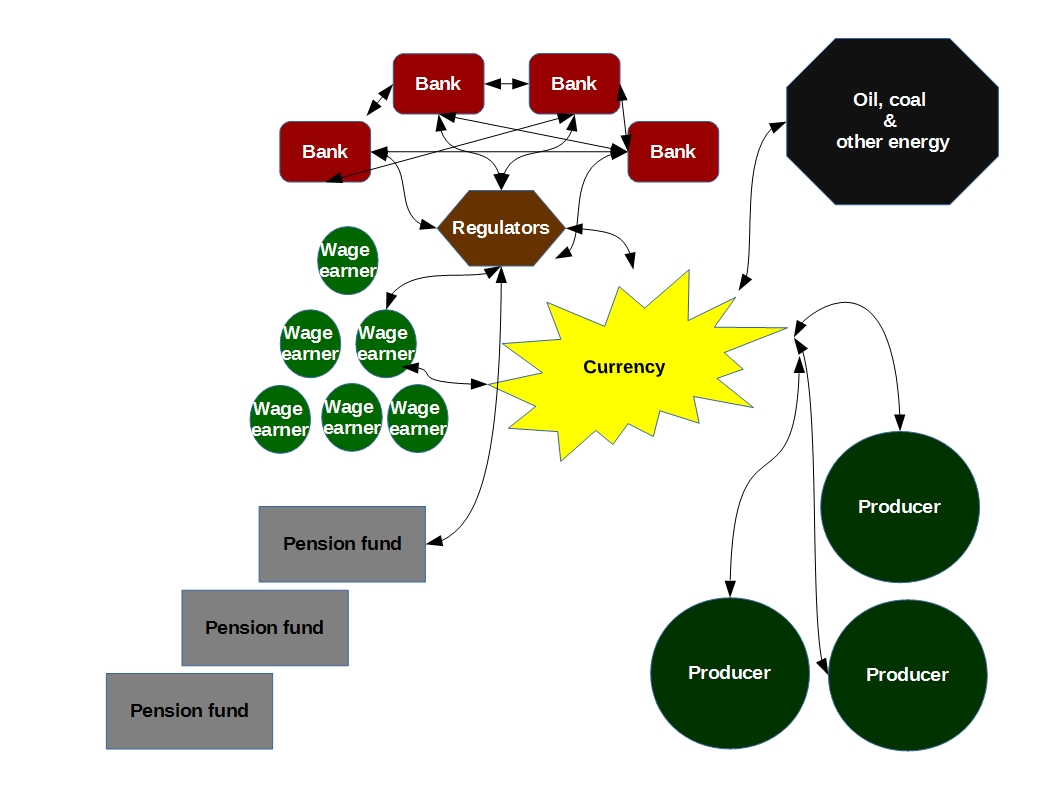}
    \caption{A simple economy with a set of key currency determinants}
\end{figure}

\begin{itemize}
    \item local variables such as GDP, employment, wages, inflation, savings and domestic interest rates
    \item capitalisation of traditional local banks and future cash-flows of their depositors
    \item differentials between local and foreign interest rates
    \item market crises due to investment bubbles or  debt accumulation in shadow banks
    \item local instability due to mismatches in expectations of various stakeholders
    \item unexpected economic frictions, for example SA electricity shortages, the impact of tax arbitrage  by multinationals  or  other endemic fraud
    \item unintended consequences of market intervention after crises in dominant markets, such as QE
    \item global dynamics, i.e. the interplay between large trading blocks, including persistent asymmetries
    \item structural changes in dominant nodes such as current developments in China
    \item unexpected economics shocks such as the knock-on effects of extreme disasters
\end{itemize}

To illustrate some dependencies, Figures 1-6 depict an  increasing complexity of modelling currency. While these simplified figures {\emph ignore full  quantitative contribution of domestic or foreign money supply}, they highlight sources of global uncertainty for emerging markets.

It follows from the interdependencies between different economies that modelling schemes which incorporate top-down (exogenous)  and bottom-up (endogenous) sources of causation offer longer-term solutions to currency-exchange valuation.

\begin{figure}
    \includegraphics[width=9cm]{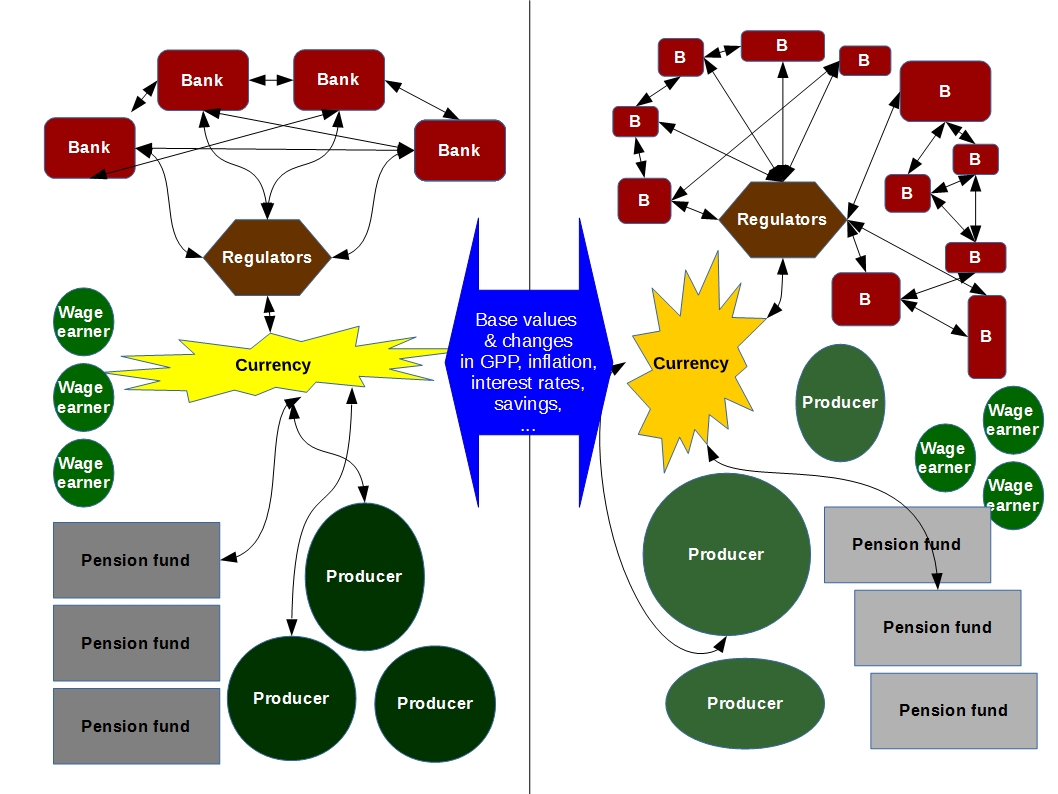}
    \caption{Two simple economies with key currency determinants. Banking systems may vary from a network of few large institutions to one comprised of several hundreds of banks.}
\end{figure}

\begin{figure}
    \includegraphics[width=9cm]{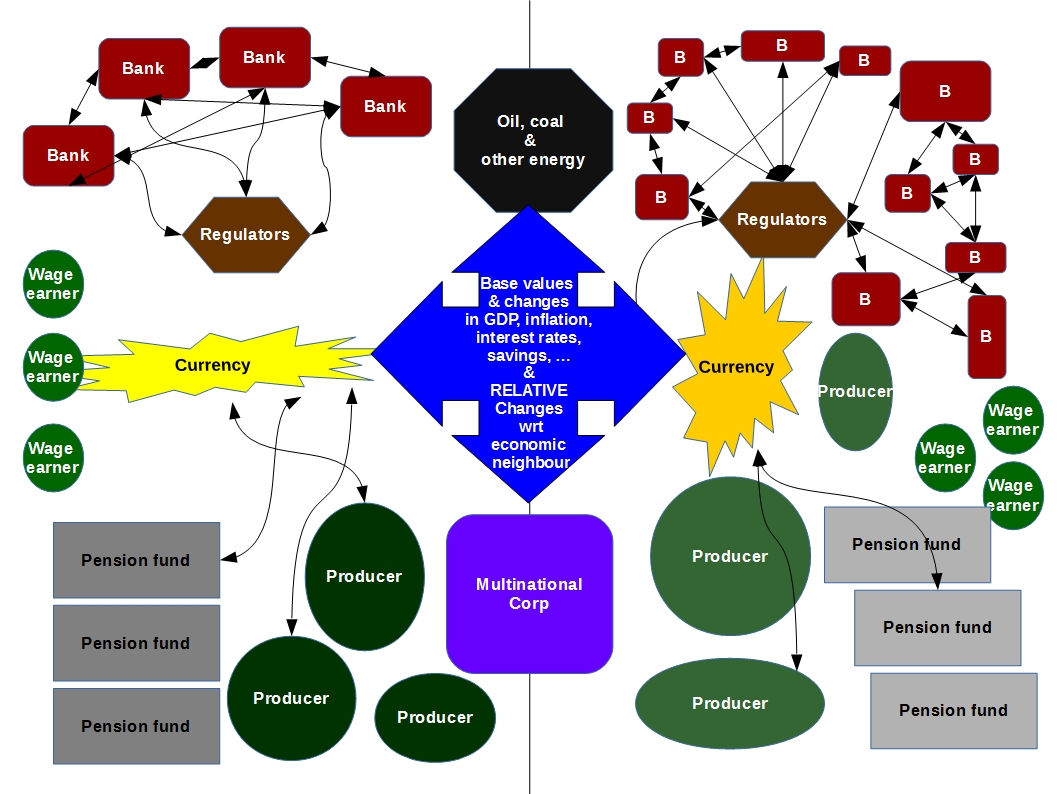}
    \caption{A relatively simple economy with key currency determinants and including energy costs and the presence of multinationals}
\end{figure}

\begin{figure}\label{fig4}
    \includegraphics[width=9cm]{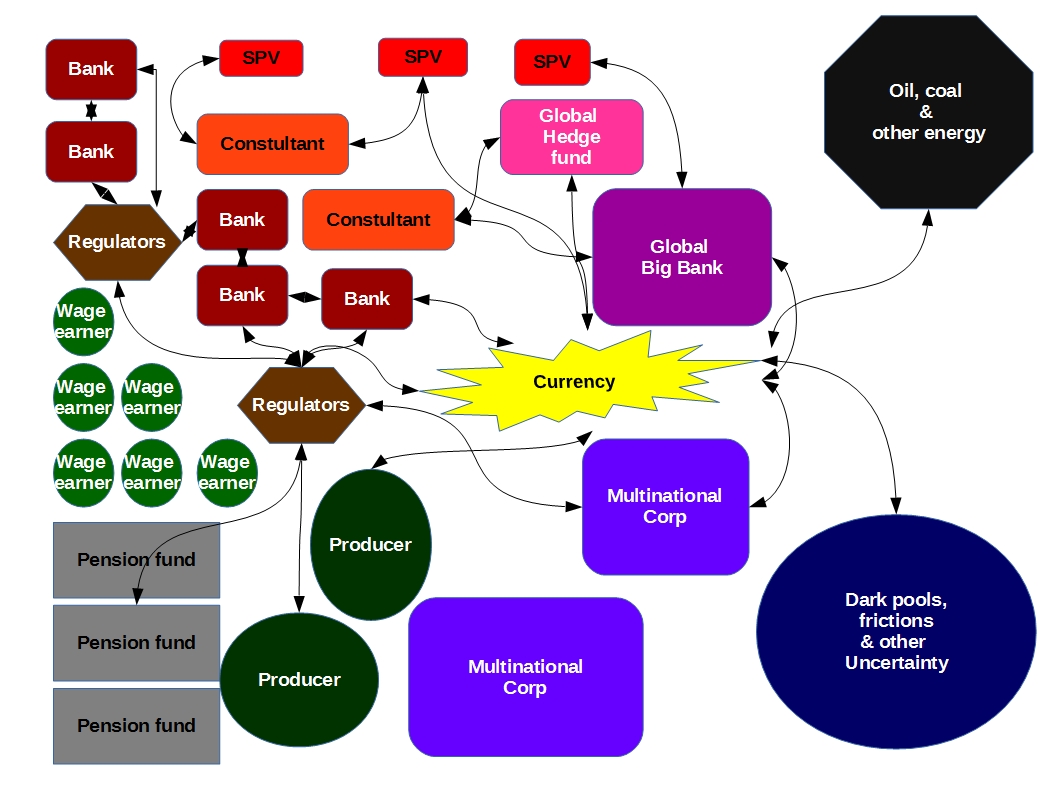}
        \caption{A more complex economy with key currency determinants, including energy costs, the presence of multinationals and global banks, dark pools, special purpose vehicles and other uncertainties.}
\end{figure}

\begin{figure}
    \includegraphics[width=9cm]{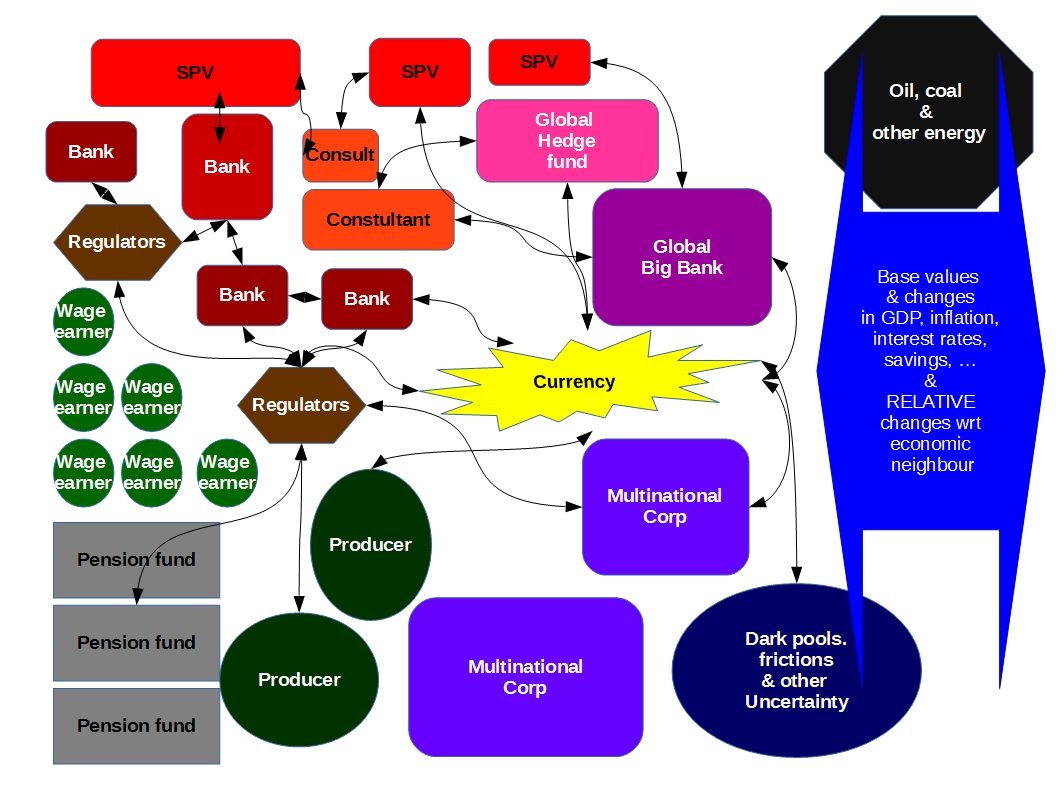}
        \caption{The same model as in Figure 4, but with a potentially unstable banking sector (top left corner), where a bank is coupled to an offspring special purpose investment vehicle whose balance sheet is bigger than the parent bank -  this scenario occurred during the credit crisis for at least one large multinational bank.}
\end{figure}

\begin{figure}
    \includegraphics[width=9cm]{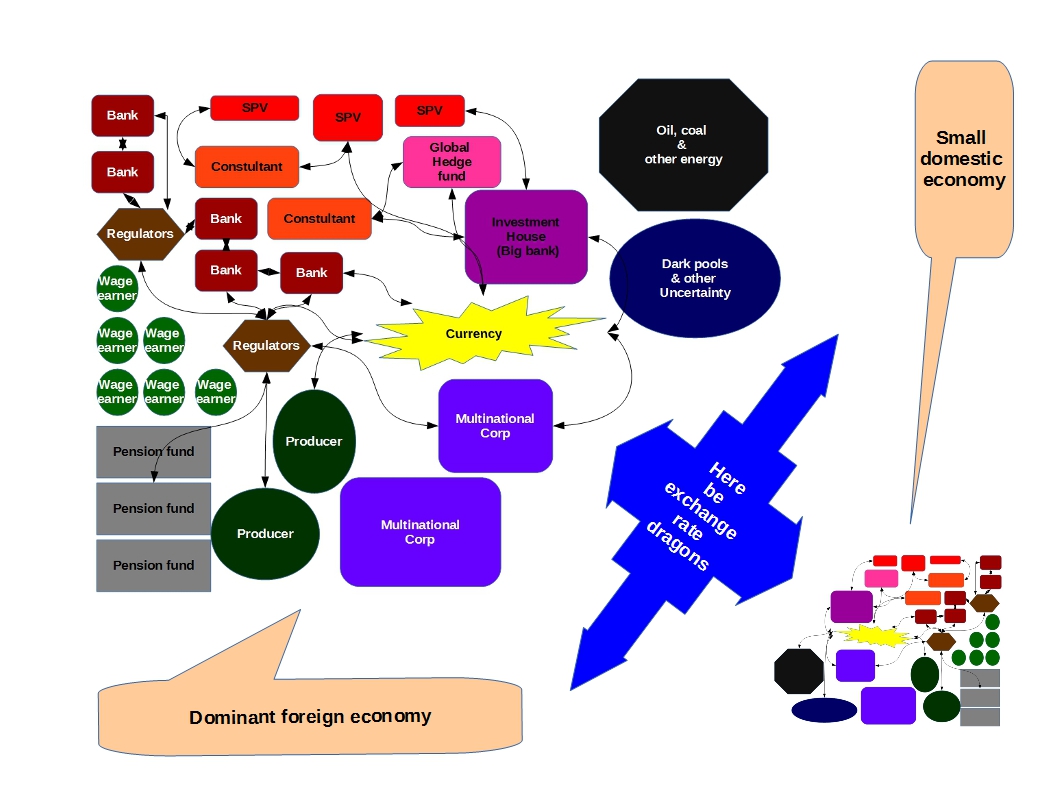}
        \caption{The same model as in Figure 4, with a complex economy coupled to a much bigger economic partner,  such that the smaller economy is impacted by internal developments in the larger.}
\end{figure}

\vspace{-0.2cm}

\section{Conclusion}

\vspace{-0.2cm}

From the previous section, exchange-rates are lynchpins which hold together various networks of transaction flows.

It is understood now that the simplified abstractions which made BSM elegant,   also led to the underpricing of risk in global markets prior to the GFC.
With changing dynamics, failures in application of even the best models become an eventual certainty.

Electronic access and algorithmic trading have provided innovative perspectives on price evolutions as markets map information about real economic data to numeric prices. If markets are free of arbitrage, then goods are probably priced efficiently.  In reality, valuation is driven at various levels of economic interaction, which are not always synchronised or equally informed. Innovations and structural transitions can have significant impact on money supply and currency valuation.  Similarly,  systemic asymmetries make smaller scale participants more vulnerable to failures, weaknesses or transitions in partner economies of larger scale.

Increased complexity demands increased sophistication and agility of regulatory oversight. Markets are neither globally free, nor globally fair. With this, comes the implication that arbitrage-free models of exchange rates are as challenging as the rewriting of economic theory itself.

\small

\vspace{1cm}
\begin{acknowledgments}
 This discussion
is based on research which has been funded in part by the National
Research Foundation of South Africa (Grant numbers 87830, 74223 and 70643).
The conclusions herein are due to the author:  any omissions or errors in reasoning are my own and should not be attributed to  my co-authors, colleagues or informal research contacts. In particular, the NRF and the University of the Witwatersrand accept no liability in this regard.
\end{acknowledgments}

\end{document}